\documentclass[11pt,twoside]{article}
\usepackage{./asp2014}

\aspSuppressVolSlug
\resetcounters

\begin{document}

\title{Evolution of intermediate mass and massive binary stars: physics, mass loss, and rotation}
\author{Dany Vanbeveren$^1$ and Nicki Mennekens$^1$
\affil{$^1$Astronomy and Astrophysics Research Group, Vrije Universiteit Brussel, Brussels, Belgium; \email{dvbevere@vub.ac.be}}}

\paperauthor{Dany Vanbeveren}{dvbevere@vub.ac.be}{}{Vrije Universiteit Brussel}{Astronomy and Astrophysics Research Group}{Brussels}{Brussels}{1050}{Belgium}
\paperauthor{Nicki Mennekens}{nmenneke@vub.ac.be}{}{Vrije Universiteit Brussel}{Astronomy and Astrophysics Research Group}{Brussels}{Brussels}{1050}{Belgium}

\begin{abstract}
In the present review we discuss the past and present status of the interacting OB-type binary frequency. We critically examine the popular idea that Be-stars and supergiant sgB[e] stars are binary evolutionary products. The effects of rotation on stellar evolution in general, stellar population studies in particular, and the link with binaries will be evaluated. Finally a discussion is presented of massive double compact star binary mergers as possible major sites of chemical enrichment of r-process elements and as the origin of recent aLIGO GW events.
\end{abstract}

\section*{Introduction}

The more and the better the observations, the more stars are found to be member of a binary. \emph{In the present paper we only consider interacting binaries} (i.e., binary = interacting binary) meaning that during the evolution of the binary at least one binary component will fill its Roche lobe. For O-B type binaries with circular orbit this means that the orbital period P should be less than 10 years (orbital separation A less than 20 AU). However, due to the eruptive mass loss in Luminous Blue Variables (LBVs like Eta Car) or due to the large stellar wind mass loss in Red Supergiants (RSGs) these P-A-limits may be considerably smaller than the values quoted above: the LBV scenario\footnote{The LBV scenario applies for binaries with at least one component with initial mass larger than 40 M$_{\odot}$} and the RSG scenario\footnote{The RSG scenario applies for binaries with at least one component with initial mass between $\sim$15-20 M$_{\odot}$ and 40 M$_{\odot}$.} of massive binaries (Vanbeveren, 1993; Vanbeveren et al., 1998b; Mennekens and Vanbeveren, 2014).  

We caution that in binaries where the LBV and RSG scenarios do not apply, the P-A-limits mentioned above may be considerably larger in case of eccentric orbits.

\section{The Galactic primordial OB-type binary frequency}

Let us first remark that a statistically significant number of OB-type binary observations is available for the Solar neighborhood only so that conclusions which are based on these observations apply for that region.  Whether these conclusions apply for the whole Galaxy, for the whole cosmos is a matter of faith.

\subsection{The status before the year 2000}

\subsubsection{The primordial O-type binary frequency}

By carefully selecting a sample of $\sim$60 O-type stars, Garmany et al. (1980) concluded that $\sim$33\% may be primary of a binary with mass ratio q > 0.2 (q = M$_\mathrm{secondary}$/M$_\mathrm{primary}$) and orbital period P < 100 days. However it was argued by Mason et al. (1998) that it is probable that many more O-type binaries with periods larger than 100 days await to be discovered.

At the first massive star conference in 1971 in Buenas Aires, Kuhi (1973) presented a statistical study showing that it is probable that all WR stars are binary components. However, Vanbeveren and Conti (1980) demonstrated that the Kuhi statistics is biased, and that the real WR+OB frequency is no more than 40\%, a binary percentage that has not changed very much over the years till now.

To estimate the primordial massive O-type binary frequency, Vanbeveren et al. (1998a) proceeded as follows. A population of massive stars consists of real single stars, un-evolved (=pre-RLOF\footnote{RLOF = Roche lobe overflow}) binaries, evolved (=post-RLOF) binaries, post-supernova rejuvenated binary mass gainers (O+compact companion or disrupted single stars but with binary origin), binary mergers (also single stars but with binary origin), etc. This means that the binary frequency in a region where star formation is continuous in time is smaller than the binary frequency at birth (=primordial binary frequency). Using a detailed binary population code Vanbeveren et al. tried to answer the question: what must be the primordial massive binary frequency f in order to explain the observed results of Garmany et al. and the observed WR binary frequency? The final answer depends on the details of all processes that govern binary evolution but with standard assumptions we concluded that f > 0.7.
	
\subsubsection{The primordial B-type binary frequency}

Abt et al. (1990) studied a sample of 109 B2-B5 stars and concluded that 29\% are spectroscopic binaries. Wolff (1978) considered 83 late B-type stars and argued that 24\% are binary with mass ratio q > 0.1 and period P < 100 days. The Bright Star Catalogue (Hoffleit and Warren, 1991) contains 511 B0-B3 stars and after carefully accounting for observational biases Vanbeveren et al. (1998b) promoted a binary frequency $\sim$35\%. Similarly as the O-type star population, the B-type star population consists of pre- and post RLOF stars, real singles but also singles with a binary origin. A similar binary population synthesis exercise as described in Vanbeveren et al. (1998a) for the O-type stars then reveals that in order to explain the $\sim$35\% binary frequency quoted above, the primordial B-type interacting binary frequency should be $\geq$ 50\%.

\subsection{The status after the year 2000}

A homogeneous sample of 71 O-type stars in 6 Galactic open clusters has been discussed by Sana et al. (2012); 40 (e.g., 40/71 = 56\%) are confirmed spectroscopic binaries. After correction for observational biases the authors conclude that the intrinsic (=primordial) O-type binary frequency $\sim$69\% which is a nice confirmation of the status before the year 2000. Most interestingly, the resulting mass ratio distribution seems to be flat whereas the period distribution is slightly skewed towards small values of P.

A radial velocity spectroscopic survey of 250 O-type stars and 540 B-type stars was presented by Chini et al. (2012). If all radial velocity variations are interpreted as being due to binarism, the authors plotted the resulting binary frequencies as function of spectral type (Fig. 3 in Chini et al.). Using the O-type frequencies depicted in that figure and performing a similar binary population synthesis experiment as discussed in section 1.1.1, one arrives at the conclusion that it can not be excluded that all O-type stars are born in binaries. The same is true for the early B-type stars. By considering the whole B-type star set, we obtain an overall B-type primordial binary frequency of 50\% supporting the status before 2000.
	
\section{The extra-Galactic binary frequency}

The effect of binaries on population synthesis of SN II and SN Ib/c supernovae has been the subject of numerous studies (Tutukov et al. 1992; Podsiadlowski et al., 1992; Joss et al., 1992; De Donder and Vanbeveren, 1998, 2003, 2004; Belczynski et al., 2002; ...some more recent papers essentially confirm the earlier ones). It was realized that the number ratio SN II rate/SN Ib/c rate depends critically on the massive binary frequency. This means that number ratio differences between different types of galaxies may reflect differences in the population of these massive binaries. De Donder and Vanbeveren (1998) noted that the data available till 1998 revealed a significant ratio difference between early type and late type galaxies, a difference which could indicate a significant difference in binary population. Obviously this 1998 research has to be reconsidered using more recent data. De Donder and Vanbeveren (1998) also compared the overall (cosmological?) observed ratio with population number synthesis predictions and concluded that the overall cosmological massive interacting binary frequency $\sim$50\%.

\section{The Be-star - binary connection}

The Be-star definition contains a list of properties but for evolutionary purposes the most important one is that Be-stars rotate at or close to the critical speed. Figure 1 illustrates the typical evolution of a binary explaining systems like $\phi$ Per or Be-type X-ray binaries. An extended description of Figure 1 is given in Vanbeveren et al. (1998c) and is not repeated here. Most importantly for the scope of the present paper is the fact that RLOF and mass accretion is accompanied by the transfer of angular momentum and the gainer spins-up. The popular idea is that such a spun-up gainer will be observed as a Be-star. During the RLOF of the binary (second peanut in Figure 1), some binaries may merge and this may obviously also result in a rapid rotator, may be a Be-star. Using this popular model it is possible to perform binary population synthesis in order to estimate the number of Be-stars formed this way, a number that can be compared to observations. This has been done by Pols et al. (1991), Van Bever and Vanbeveren (1997) and repeated with some updated physics in the PhD-thesis of Joris Van Bever (2004). Using an initial binary frequency of 70\% the binary population number synthesis simulations of Van Bever (2004) are illustrated in Figure 2 showing the predicted Be-star frequency as function of mass. The dotted lines indicate (within the uncertainties) the observations as discussed by Zorec and Briot (1997). It is tempting then to conclude that it cannot be excluded that most (all) Be-stars are formed by binary mass exchange and/or binary mergers. There are however a number of counter arguments.

\articlefigure[width=.58\textwidth]{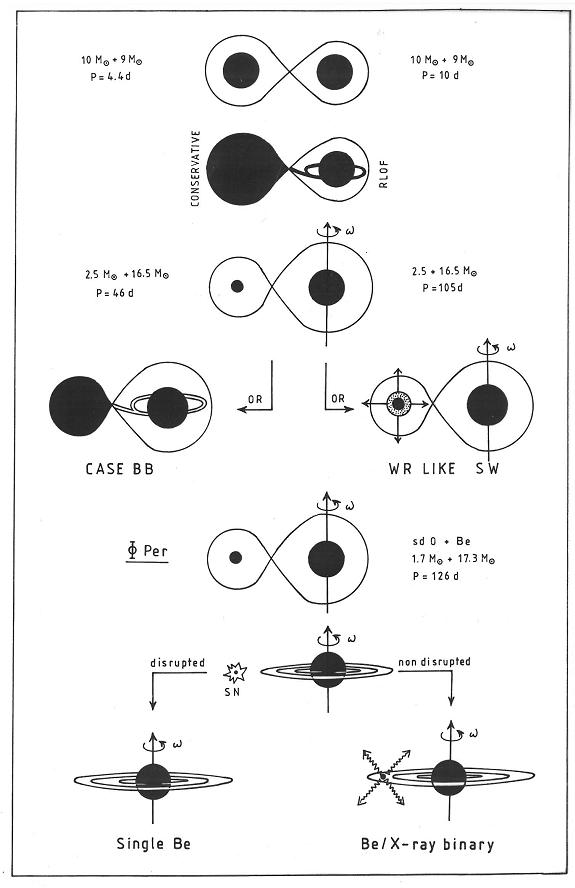}{fig1}{Typical evolution of a binary explaining systems like $\phi$ Per or Be-type X-ray binaries (from Vanbeveren et al., 1998c).}

\articlefigure[width=.64\textwidth]{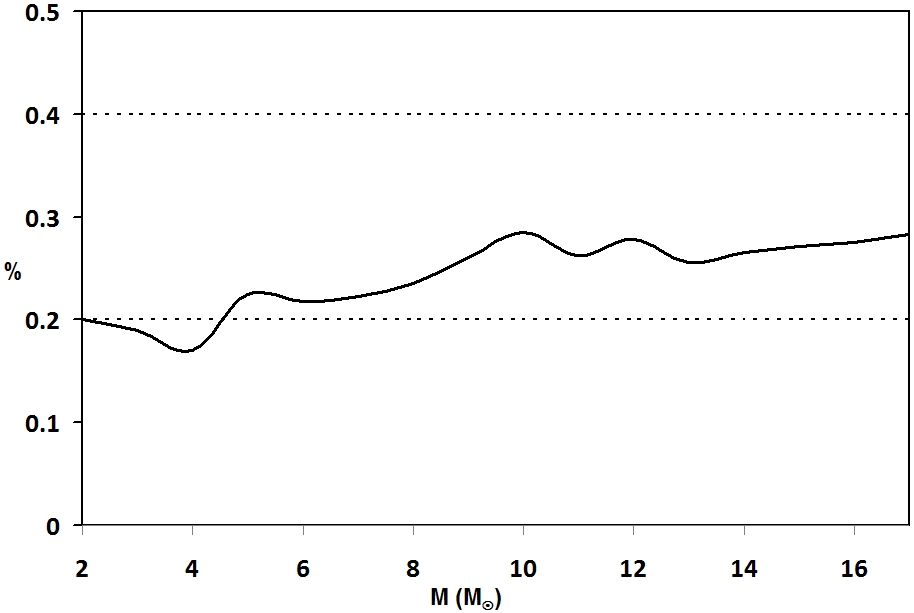}{fig2}{Be-star frequency as a function of mass. Solid: predicted; dotted lines: min-max observations of Zorec and Briot (1997).}

\begin{itemize}
\item The mass gainers in most large period\footnote{Here large means that the action of tidal synchronization is small; in the case of Algols this means periods larger than $\sim$10 days.} Algols rotate at 10-40\% of their critical speed (Dervisoglu et al., 2010) with some exceptions (RW Per, RX Sct). The authors argue that there must be an efficient mechanism for angular momentum loss from accretors and they propose accretion/rotation induced magnetic fields.
\item A statistically significant number of Be stars are found in binaries with orbital separation larger than 20 AU (Oudmaijer and Parr, 2010) and this seems to rule out binary mass exchange as a cause for the Be-star formation.
\item If most of the Be stars would be spun-up binary mass gainers or binary mergers, we would expect that most of the Be-stars in clusters are rejuvenated blue stragglers. This is illustrated in Van Bever \& Vanbeveren (1997, their fig. 3) where the typical evolution of the HR-diagram of a cluster is shown with a significant number of binaries at birth (see also Van Bever, 2004 for an update). However, this is not supported by observations.
\item Last but not least, Table 1 gives the age and the observed Be fraction of a number of well populated clusters in the Galaxy and the Magellanic Clouds and these numbers are compared to theoretical prediction if indeed Be stars would be spun-up mass gainers or binary mergers and assuming a primordial cluster binary frequency of 70\%. As can be noticed the predicted values are up to a factor 10 lower than observed.
\end{itemize}

\begin{table}[!ht]
\caption{Age and observed Be fraction of a number of well populated clusters in the Galaxy and the Magellanic Clouds, compared to theoretical prediction.}
\smallskip
\begin{center}
{\small
\begin{tabular}{lllll}
\tableline
\noalign{\smallskip}
 &  & Age (Myr) & Observed & Predicted B-type mass\\
 &  &  & Be fraction & gainer fraction\\
\noalign{\smallskip}
\tableline
\noalign{\smallskip}
Galaxy & $\eta$ \& $\chi$ Per & < 20 & 0.2-0.5 & 0.02-0.04 \\
 & NGC 663 & 22 & 0.4 & 0.03-0.05 \\
 & NGC 3760 & 22 & 0.33 & 0.03-0.05 \\
\noalign{\smallskip}
MCs & NGC 330 & 19 & 0.27 & 0.02-0.04 \\
 & NGC 2004 & 20 & 0.11 & 0.02-0.04 \\
 & NGC 1818 & 25 & 0.2 & 0.04-0.05 \\
\noalign{\smallskip}
\tableline\
\end{tabular}
}
\end{center}
\end{table}

It is clear that the origin of the Be-phenomenon (star formation and/or binary evolution) is still a matter of debate. Of course there are Be-stars in relatively short period binaries, e.g., the Be-X-ray binaries and some systems like $\phi$ Per. Compared to the popular model an alternative one could be that \emph{Be-stars in binaries are born as rapidly rotating (Be) stars}. As an illustration: start with a 15 M$_{\odot}$ + 12 M$_{\odot}$ binary with a period of 1500 days, where the 12 M$_{\odot}$ star is a rapid rotator (Be-type). The RLOF of the 15 M$_{\odot}$ star rapidly leads to the formation of a common envelope and in combination with RSG stellar wind the system evolves into a 3 M$_{\odot}$ + 12 M$_{\odot}$ (Be-type) binary with a period $\sim$130 days (a system like $\phi$ Per). When the 3 M$_{\odot}$ explodes (eventually an electron capture SN) and the binary remains bound, the system becomes a 1.4 M$_{\odot}$ (neutron star) + 12 M$_{\odot}$ Be (Be-X-ray binary). Note that MWC 656 is a Black Hole + Be binary candidate and also here it has been suggested that the Be-star is born as rapid rotator (Grudzinska et al., 2015).

\section{Massive binary mergers and the supergiant sgB[e] connection}

If sgB[e] stars are rapid rotators like the Be-star cousins, the question rises how to make a rapidly rotating B-type supergiant? Vanbeveren et al. (2013) and Justham et al. (2014) proposed that B[e] stars may be case Br\footnote{Case Br means that the period of the binary is such that RLOF will start while the primary (= mass loser) is a hydrogen shell burning star with an envelope in radiative equilibrium.} binary mergers and quantitative computations were presented in both papers. Note that such a merger may explode as a blue supergiant.

\section{What WR+O binaries may tell us about massive binary evolution}

Has RLOF played a significant role in the formation of WR+O binaries? This question has been considered in many papers in the past (for reviews see, e.g., Vanbeveren et al., 1998b; Langer, 2012) but a definite observational proof is still lacking. If RLOF (and mass transfer) has been important, one may expect that the O-type stars (the mass gainers) are rapid rotators. An observational campaign has been set up to use SALT to measure rotational velocities in as many WR+O binaries as possible. At present we have data for 10 systems (Shara et al., 2016) and in the majority of them \emph{the O-type star is rotating super-synchronous, strongly supporting the RLOF-mass transfer formation scenario for these binaries}.

\section{Rotation of massive stars, evolution and population synthesis}

The distribution of rotational velocities of Galactic O-type stars has been discussed by Conti \& Ebbets (1977), Penny (1996), Vanbeveren et al. (1998a) and of O-type stars in the LMC by Ramirez-Agudelo et al. (2013). Both distributions are shown in Figure 3. Notice: a. the relative similarity of both distributions, b. the fact that most O-type stars are relatively slow rotators, c. the fact that both distributions have a tail towards large velocities. A close inspection of the Galactic tail reveals that it contains most of the O-type runaways (Vanbeveren, 2009) suggesting a binary origin for these rapid rotators, much like the O-stars in WR+O binaries discussed in the previous section. de Mink et al. (2013) quantitatively verified this suggestion by using population synthesis tools. They started with a population of relatively slow rotating stars (with a uniform distribution with average = 100 km/s and standard deviation = 60 km/s) and with 70\% binaries and they showed that as a consequence of the action of RLOF, mass transfer and mergers the final rotation velocity distribution of field O-type stars very closely resembles the one of Figure 3. Moreover, this simulation illustrates that it cannot be excluded that \emph{the influence of rotation on the evolution of single stars is very moderate}.

\articlefigure[width=.68\textwidth]{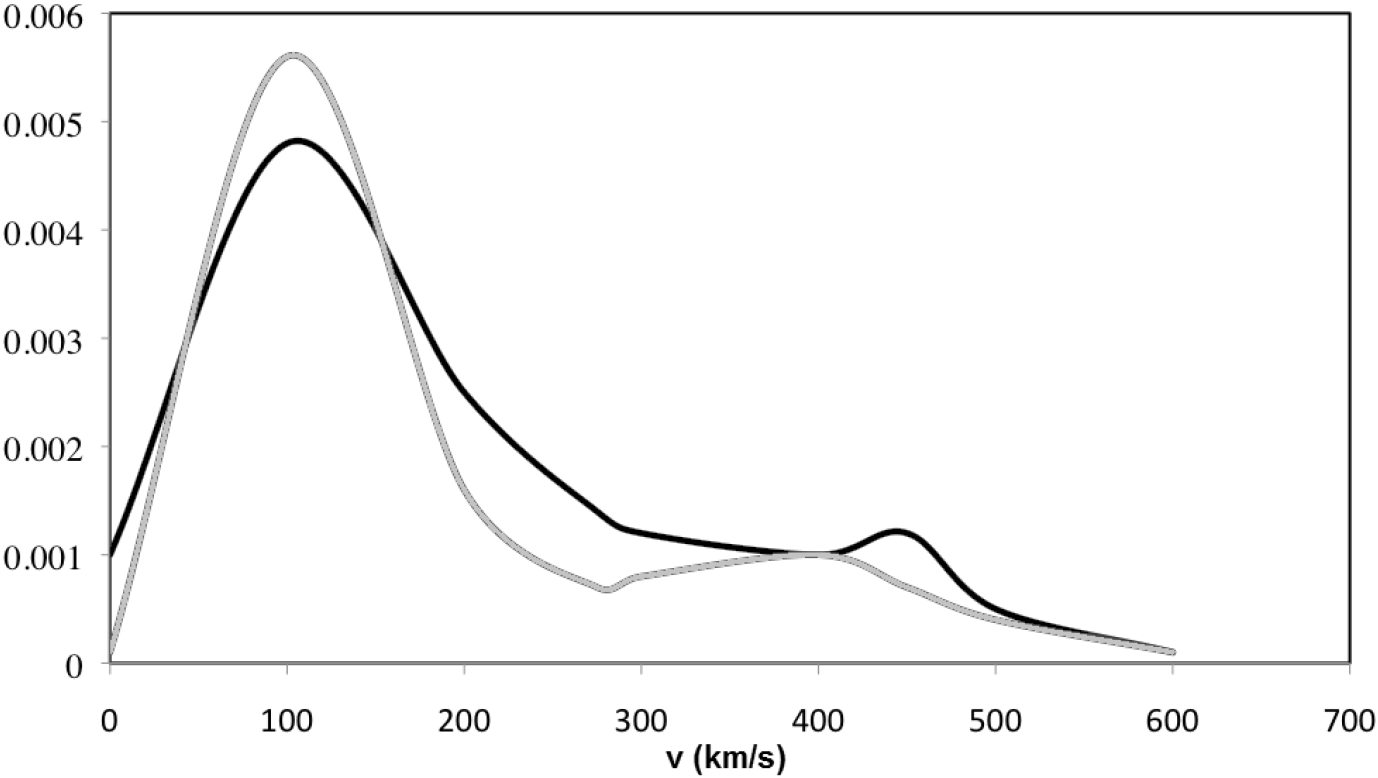}{fig3}{Distribution of rotational velocities of O-type stars. Black: LMC; gray: Galaxy.}

\section{Double compact star binary mergers: important production sites of r-process elements}

Recent hydrodynamical simulations of the merging process of double neutron star binaries (NSNS) or neutron star + black hole binaries (NSBH) have shown that during and after the merger phase some 0.001-0.1 M$_{\odot}$ of neutron rich matter can be ejected, and that this matter provides the necessary conditions for efficient r-processing (Korobkin et al., 2012; Just et al., 2015 and references therein).

By implementing the details of one particular r-process site in a chemical evolutionary code, it is possible to compute the galactic temporal variation of the r-process elements predicted by that site. Comparison with observations (mainly the observations of Eu) may yield important information about the importance of the chosen site. In principle, to study in this way the effects of merging NSNS and/or NSBH, one has to combine a full binary population model (including binary chemical yields) with a galactic formation and evolution code. Our code, described in De Donder \& Vanbeveren (2003, 2004), Vanbeveren et al. (2012) and in Mennekens \& Vanbeveren (2014, 2016), is to our knowledge the only code with such a self-consistent model. Other studies (e.g., Bauswein et al., 2014; Matteucci et al., 2014; Vangioni et al., 2016) use a galactic code that accounts for single stars only and the importance of binaries is then parameterized by using some existing delayed time distribution (DTD) for NSNS and/or NSBH mergers published elsewhere. The three papers cited above all use a DTD published by Dominic et al. (2012) and references therein. However, it has to be remarked that the latter DTD does not include the lifetime of the progenitor of the double compact star binary and this is of prime importance for the evolution of the r-process elements during the early phases of the chemical evolution of a galaxy. Moreover, DTDs of NSNS and NSBH mergers depend on a number of parameters governing binary evolution and it may therefore be advisable to work more in the line of self-consistency. Note that already in our 2003, 2004 work we concluded that \emph{NSNS/NSBH mergers are major r-process production sites; to explain the early galactic observations an additional process may be needed.}

The hydro-simulations of NSNS/NSBH mergers have reached such a stage of robustness that it may become possible to propose strong upper-limits on the rate of merger events by comparing the predicted r-process element enrichment with the observed one. Bauswein et al. (2014) proposed a strict upper limit of 60 NSBH mergers per Myr for our Galaxy whereas Mennekens \& Vanbeveren (2014) concluded that \emph{there should be no more than 20 NSNS/NSBH mergers per Myr}. These results are of course important for the interpretation of future aLIGO measurements.

\section{Some interesting test beds}

\subsection{GW150914}

Till now, aLIGO has detected three gravitational wave events that most probably resulted from the merging of two black holes: GW 150914, LVT 151012 and GW 151226. The first one is the most suggestive since it consisted of two very massive black holes with masses 36$\pm$5 M$_{\odot}$ and 29$\pm$4 M$_{\odot}$. First note that the progenitors of such massive BHs may be stars with very large initial mass, significantly larger than 40 M$_{\odot}$. In literature three formation-scenarios have been worked out in some detail:

\begin{itemize}
\item The very massive binary evolution model in the field, in a region with small metallicity and via common envelope evolution (Dominic et al., 2012; Belczynski et al., 2016)
\item The formation of very massive binary BHs in Globular Clusters (Rodriguez et al., 2016) due to dynamical processes in dense stellar regions
\item Chemically homogeneous evolution of very tight massive binaries (early Case A) (Marchant et al., 2016).
\end{itemize}

The first one has been criticized by Mennekens \& Vanbeveren (2014). The region in the HR-diagram above 40 M$_{\odot}$ is occupied by Luminous Blue Variables (LBVs) and does not contain Red Supergiants. Even more, the masses of GW 150914 suspect a progenitor binary with component masses similar to the progenitor mass of $\eta$ Car. LBVs are characterized by eruptions where the star loses a lot of mass on a very short timescale, and high and more or less stable stellar wind mass loss in between the eruptions. The physical process of the eruptions is still unknown but we argued that within our present knowledge it cannot be excluded that due to the combined effect of LBV eruptions and LBV stellar wind mass loss the common envelope phase is avoided in very massive binaries (the LBV scenario of massive binaries, see also the introduction). Contrary to the common envelope process, LBV mass loss significantly increases the binary period and it was shown by Mennekens \& Vanbeveren (2014) that in this case the resulting binary BH has such a large period that merging within the Hubble time does not happen. A critical assumption here is of course that the LBV phenomenon in general, the LBV eruptions in particular, also happens in low metallicity regions. Note that we questioned the first formation-scenario. If our criticism is valid, it would make the other formation-scenarios more plausible.

\subsection{MWC 656}

MWC 656 is a BH+Be binary candidate (Casares et al., 2014). Grudzinska et al. (2015) propose a binary evolutionary model where the Be-star is born as a Be-star. They start with a binary with period > 1000 days, with primary mass larger than 40 M$_{\odot}$ and a 13 M$_{\odot}$ secondary born as rapid rotator. The system is assumed to evolve through a common envelope phase finally forming a 60 day period binary with a > 5 M$_{\odot}$ BH and a 13 M$_{\odot}$ Be star. 
Accounting for the LBV scenario discussed in the previous subsection we propose an alternative. We also start with the same primary and secondary as in the model above but the initial binary period is 15 days only. Due to LBV eruptions/stellar wind and WR winds during core helium burning, the common envelope/RLOF is avoided. LBV eruptions/stellar wind and WR wind causes the binary period to increase and the final outcome is a system like MWC 656.

\subsection{The high mass X-ray binary Vela X-1}

As discussed by Vanbeveren et al. (1994) the optical star in Vela X-1 is over-luminous and this may indicate that during the mass transfer phase of the progenitor binary, the mass gainer was efficiently mixed, possibly as a consequence of the spin-up process that accompanies mass transfer during RLOF. This system certainly deserves more investigation.

\subsection{$\zeta$ Pup}

$\zeta$ Pup is the brightest O-type star in the sky. Since it is (till now) single, it has for long been considered as a gift from the gods as far as single star evolution is concerned. However, the star is a runaway (a fact which has been overlooked for quite some time in the past), which means that perhaps it is a gift from the gods as far as binaries are concerned. To explain the runaway nature two binary models have been explored: a model based on the binary supernova model of Blaauw (1961) (Vanbeveren et al., 1998) and the dynamical encounter of two objects (single-binary or binary-binary) in the core of a dense cluster (Vanbeveren, 2012).

\section{Conclusions}

We propose the following conclusions that deserve further investigation:

\begin{itemize}
\item The interacting OB-binary frequency is > 50-70\%. This means that population studies where binaries are ignored may have an academic value but may be far from reality
\item At least 20\% of the B-type star population are binary mergers or binary mass gainers but not all are Be-type stars; may be only few Be-type stars are formed via binary interaction
\item sgB[e]-stars may be case Br binary mergers
\item aLIGO-rates in general, the formation of double BH mergers in particular, depend critically on the physics of the LBV-phenomenon.
\end{itemize}


\begin{thebibliography}{}
\bibitem[]{} Abt H., Gomez A. \& Levi S. 1990, ApJS, 74, 551
\bibitem[]{} Bauswein, A., Ardevol Pulpillo, R., Janka, H. et al. 2014, ApJ, 795, L9
\bibitem[]{} Belczynski, K., Kalogera, V. \& Bulik, T. 2002, ApJ, 572, 407
\bibitem[]{} Belczynski, K., Holz, D., Bulik, T. et al. 2016, Nature, 534, 512
\bibitem[]{} Blaauw, A. 1961, Bull. Astr. Inst. Netherlands, 15, 265
\bibitem[]{} Casares, J., Negueruela, I., Rib\'o, M. et al. 2014, Nature, 505, 378
\bibitem[]{} Chini, R., Hoffmeister, V., Nasseri, A. et al. 2012, MNRAS, 424, 1925
\bibitem[]{} Conti, P. \& Ebbets, D. 1977, ApJ, 213, 438
\bibitem[]{} De Donder, E. \& Vanbeveren, D. 1998, A\&A, 333, 557
\bibitem[]{} De Donder, E. \& Vanbeveren, D. 2003, NewA, 8, 415
\bibitem[]{} De Donder, E. \& Vanbeveren, D. 2004, NewAR, 48, 861
\bibitem[]{} de Mink, S., Langer, N., Izzard, R. et al. 2013, ApJ, 764, 166
\bibitem[]{} Dervisoglu, A., Tout, C. \& Ibanoglu, C. 2010, MNRAS, 406, 1071
\bibitem[]{} Dominik, M., Belczynski, K., Fryer, C. et al. 2012, ApJ, 759, 52
\bibitem[]{} Garmany, C., Conti, P. \& Massey, P. 1980, ApJ, 242, 1063
\bibitem[]{} Grudzinska, M., Belczynski, K., Casares, J. et al. 2015, MNRAS, 452, 2773
\bibitem[]{} Hoffleit, D. \& Warren, W. 1991, The Bright Star Catalogue, 5th Revised Ed., Yale University Observatory
\bibitem[]{} Joss, P., Hsu, J., Podsiadlowski, P. et al. 1992, IAUS, 151, 523
\bibitem[]{} Just, O., Bauswein, A., Ardevol Pulpillo, R. et al. 2015, MNRAS, 448, 541
\bibitem[]{} Justham, S., Podsiadlowski, P. \& Vink, J. 2014, ApJ, 796, 121
\bibitem[]{} Korobkin, O., Rosswog, S., Arcones, A. et al. 2012, MNRAS, 426, 1940
\bibitem[]{} Kuhi, L. 1973, IAUS, 49, 205
\bibitem[]{} Langer, N. 2012, ARA\&A, 50, 107
\bibitem[]{} Marchant, P., Langer, N., Podsiadlowski, P., et al., 2016, A\&A, 588, A50
\bibitem[]{} Mason, B., Gies, D., Hartkopf, W. et al. 1998, AJ, 115, 821
\bibitem[]{} Matteucci, F., Romano, D., Arcones, A. et al. 2014, MNRAS, 438, 2177
\bibitem[]{} Mennekens, N. \& Vanbeveren, D. 2014, A\&A, 564, A134
\bibitem[]{} Mennekens, N. \& Vanbeveren, D. 2016, A\&A, 589, A64
\bibitem[]{} Oudmaijer, R. \& Parr, A. 2010, MNRAS, 405, 2439
\bibitem[]{} Penny, L. 1996, ApJ, 463, 737
\bibitem[]{} Podsiadlowski, P., Joss, P. \& Hsu, J. 1992, ApJ, 391, 246
\bibitem[]{} Pols, O., Cot\'e, J., Waters, L. \& Heise, J. 1991, A\&A, 241, 419
\bibitem[]{} Ramirez-Agudelo, O., Simon-Diaz, S., Sana, H. et al. 2013, A\&A, 560, A29
\bibitem[]{} Rodriguez, C., Chatterjee, S. \& Rasio, F. 2016, Phys Rev D, 93, 084029
\bibitem[]{} Sana, H., de Mink, S., de Koter, A. et al. 2012, Sci, 337, 444
\bibitem[]{} Shara, M., Crawford, S., Vanbeveren, D. et al. 2016, arXiv:1511.00046
\bibitem[]{} Tutukov, A., Yungelson, L. \& Iben, I. 1992, ApJ, 386, 197
\bibitem[]{} Van Bever, J. 2004, PhD thesis, Vrije Universiteit Brussel
\bibitem[]{} Van Bever, J. \& Vanbeveren, D. 1997, A\&A, 322, 116
\bibitem[]{} Vanbeveren, D. 1993, SSRv, 66, 327
\bibitem[]{} Vanbeveren, D. 2009, NewAR, 53, 27
\bibitem[]{} Vanbeveren, D., 2012, ASPC, 465, 342
\bibitem[]{} Vanbeveren, D. \& Conti, P. 1980, A\&A, 88, 230
\bibitem[]{} Vanbeveren, D., Mennekens, N., Van Rensbergen, W. et al. 2013, A\&A, 552, A105
\bibitem[]{} Vanbeveren, D., Herrero, A., Kunze. D. et al. 1994, SSRv, 66, 395
\bibitem[]{} Vanbeveren, D., De Donder, E., Van Bever, J. et al. 1998a, NewA, 3, 443
\bibitem[]{} Vanbeveren, D., Van Rensbergen, W. \& De Loore, C. 1998b, A\&AR, 9, 63
\bibitem[]{} Vanbeveren, D., Van Rensbergen, W. \& De Loore, C. 1998c, The Brightest Binaries, Kluwer Academic Publishers
\bibitem[]{} Vangioni, E., Goriely, S., Daigne, F. et al. 2016, MNRAS, 455, 17
\bibitem[]{} Wolff, S. 1978, ApJ, 222, 556
\bibitem[]{} Zorec, J. \& Briot, D. 1997, A\&A, 318, 443
\end{thebibliography}
\end{document}